# NONSEPARABILITY, POTENTIALITY AND THE CONTEXT-DEPENDENCE OF QUANTUM OBJECTS

VASSILIOS KARAKOSTAS*

SUMMARY. Standard quantum mechanics undeniably violates the notion of separability that classical physics accustomed us to consider as valid. By relating the phenomenon of quantum nonseparability to the all-important concept of potentiality, we effectively provide a coherent picture of the puzzling entangled correlations among spatially separated systems. We further argue that the generalized phenomenon of quantum nonseparability implies contextuality for the production of well-defined events in the quantum domain, whereas contextuality entails in turn a structural-relational conception of quantal objects, viewed as carriers of dispositional properties. It is finally suggested that contextuality, if considered as a conditionalization preparation procedure of the object to be measured, naturally leads to a separable concept of reality whose elements are experienced as distinct, well-localized objects having determinate properties. In this connection, we find it necessary to distinguish the meaning of the term reality from the criterion of reality for us. The implications of the latter considerations for the notion of objectivity in quantum mechanics are also discussed.

1. THE MEANING OF THE SEPARABILITY PRINCIPLE IN CLASSICAL PHYSICS

Classical physics is essentially atomistic in character. It portrays a view of the world in terms of analyzable, separately existing but interacting self-contained parts. Classical physics is also reductionistic. It aims at explaining the whole of forms of physical existence, of structures and relations of the natural world in terms of a salient set of elementary material objects linked by forces. Classical physics (and practically any experimental science) is further based on the Cartesian dualism of 'res cogitans' ('thinking substance') and 'res extensa' ('extended substance'), proclaiming a radical separation of an objective external world from the knowing subject that allows no possible intermediary.

* Department of Philosophy and History of Science, University of Athens, Athens 157 71, Greece (E-mail: karakost@phs.uoa.gr)

In fact, the whole edifice of classical physics — be it point-like analytic, statistical, or field theoretic — is compatible with the following *separability principle* that can be expressed schematically as follows:

*Separability Principle*: The states of any spatio-temporally separated subsystems $S_1$, $S_2$, ..., $S_N$ of a compound system S are individually well-defined and the states of the compound system are wholly and completely determined by them and their physical interactions including their spatio-temporal relations (cf. Howard 1989; Healey 1991).

In the case, for instance, of point-like analytic mechanics, the state of a compound system consisting of N point particles is specified by considering all pairs $\{q_{3N}(t), p_{3N}(t)\}$ of the generalized position and momentum coordinates of the individual particles. Hence, at any temporal moment t, the individual pure state of the compound system consists of the N-tuple $\omega = (\omega_1, \omega_2, ... , \omega_N)$, where $\{\omega_i\} = \{q_i, p_i\}$ are the pure states of its constituent subsystems. It is then clear that in the individual, analytical interpretation of classical mechanics maximal knowledge of the constituent parts of a compound system provides maximal knowledge of the whole system (see, for example, Scheibe 1973, pp. 53-54). Accordingly, every property the compound system has at time t, if encoded in $\omega$, is determined by $\{\omega_i\}$. For instance, any classical physical quantities (such as mass, momentum, angular momentum, kinetic energy, center of mass motion, gravitational potential energy, etc.) pertaining to the overall system are determined in terms of the corresponding quantities of its parts. They either constitute direct sums or ordinary functional relations (whose values are well-specified at each space-time point) of the relevant quantities of the subsystems. Thus, they are wholly determined by the subsystem states. Furthermore, given the state $\omega_t(q, p)$ of a classical system in phase space at time t, the dynamical law which connects $\omega_t$ with the state $\omega_{t'}(q, p)$ of the system at any other time t´ is given by the Hamiltonian $H(q, p)$ and the canonical equations. This means that a classical system $S_t$, uniquely defined at time t, can be re-identified at any other time $t' \neq t$ by the phase point $(p_t, q_t)$ values on its dynamical trajectory. Hence, classical physics determines objects-systems as individuals with temporal identity. They can be identified through conservation of their essential quantities, re-identified in time, and distinguished from their like. The foregoing concise analysis delimits actually the fact, upon which the whole classical physics is founded, that any compound physical system of a classical universe can be conceived of as consisting of *separable*, *individual* parts interacting by means of forces, which are encoded in the Hamiltonian function of the overall system, and that, if the full Hamiltonian is known, maximal knowledge of the values of the physical quantities pertaining to each one of these parts yields an exhaustive knowledge of the whole compound system, in perfect conformity with the aforementioned separability principle.

The notion of separability has been viewed within the framework of classical physics as a principal condition of our conception of the world, a condition that characterizes all our thinking in acknowledging the physical identity of distant things, the "mutually independent existence (the 'being thus')" of spatio-temporally separated systems (Einstein 1948/1971, p. 169). The primary implicit assumption pertaining to this view is a presumed *absolute kinematic independence* between the knowing subject (the physical scientist) and the object of knowledge, or equivalently, between the measuring system (as an extension of the knowing subject) and the system under measurement. The idealization of the kinematically independent behavior of a physical system is possible in classical physics both due to the Cartesian-product structure of phase space, namely, the state-space of classical theories, and the absence of genuine indeterminism in the course of events or of an element of chance in the measurement process. During the act of measurement a classical system conserves its identity. Successive measurements of physical quantities, like position and momentum that define the state of a classical system, can be performed to any degree of accuracy and the results combined can completely determine the state of the system before and after the measurement interaction, since its effect, if not eliminable, takes place continuously in the system's state-space and is therefore predictable in principle.

Consequently, classical physical quantities or properties are taken to obey a so-called *'possessed values'* principle, in the sense that the values of classical properties are considered as being possessed by the object itself independently of any measurement act. That is, the properties possessed by an object depend in no way on the relations obtaining between it and a possible experimental context used to bring these properties about. No qualitatively new elements of reality are produced by the interaction of a classical system with the measuring apparatus. The act of measurement in classical physics is passive; it simply reveals a fact which has already occurred. In other words, a substantial distinction between *potential* and *actual* existence is rendered obsolete in classical mechanics. Within the domain of the latter, all that is potentially possible is also actually realized in the course of time, independently of any measuring interventions. It should be noted, in this respect, that this is hardly the case in the quantum theory of the measurement process.

## 2. NONSEPARABILITY AND ITS RELATION TO POTENTIALITY IN QUANTUM PHYSICS

In contrast to classical physics, standard quantum mechanics systematically violates the conception of separability.[1] From a formal point of view, the source of its defiance is due to the tensor-product structure of a compound Hilbert space, which is not simply restricted to the topological (Cartesian) state-space but it includes it as a proper subset, and the quantum-mechanical principle of the superposition of states, which incorporates a kind of objective indefiniteness for the numerical values of any observable belonging to a superposed state. As a means of explicating the preceding factors in concrete physical terms, let us consider the simplest possible case of a compound system S consisting of a pair of subsystems $S_1$ and $S_2$ with corresponding Hilbert spaces $H_1$ and $H_2$. Naturally, subsystems $S_1$ and $S_2$, in forming system S, have interacted by means of forces at some time $t_0$ and suppose that at times $t > t_0$ they are spatially separated. Then, any pure state W of the compound system S can be expressed in the tensor-product Hilbert space $H = H_1 \otimes H_2$ in the Schmidt form

$$W = P_{|\Psi\rangle} = |\Psi\rangle\langle\Psi| = \sum_i c_i (|\psi_i\rangle \otimes |\varphi_i\rangle), \qquad \| |\Psi\rangle \|^2 = \sum_i |c_i|^2 = 1 , \qquad (1)$$

where $\{|\psi_i\rangle\}$ and $\{|\varphi_i\rangle\}$ are orthonormal vector bases in $H_1$ (of $S_1$) and $H_2$ (of $S_2$), respectively.

If there is just one term in the W-representation of Eq. (1), i.e., if $|c_i| = 1$, the state $W = |\psi\rangle \otimes |\varphi\rangle$ of the compound system forms a product state: a state that can always be decomposed into a single tensor-product of an $S_1$-state and an $S_2$-state. In this circumstance, each subsystem of the compound system possesses a separable and well-defined state, so that the state of the overall system consists of nothing but the sum of the subsystem states in consonance with the separability principle of Section 1. This is the only highly particular as well as idealised case in which a separability principle holds in quantum mechanics. For, even if a compound system at a given temporal instant is appropriately described by a product state — $W(t) = |\psi_{(t)}\rangle \otimes |\varphi_{(t)}\rangle$ — the preservation of its identity under the system's natural time evolution — $W(t_2) = U(t_2-t_1) W(t_1)$, for all $t \in R$ — implies that the Hamiltonian $H$ (i.e., the energy operator of the system) should be

decomposed into the direct sum of the subsystem Hamiltonians — $H = H_1 \otimes I_2 + I_1 \otimes H_2$ — and this is precisely the condition of no interaction between $S_1$ and $S_2$ (e.g., Blank et al. 1994, Ch. 11). Obviously, in such a case, subsystems $S_1$ and $S_2$ behave in an entirely uncorrelated and independent manner. Correlations, even of a probabilistic nature, among any physical quantities corresponding to the two subsystems are simply non existent, since for any two observables $A_1$ and $A_2$ pertaining to $S_1$ and $S_2$, respectively, the probability distributions of $A_1$ and of $A_2$ are disconnected: Tr $(A_1 \otimes A_2)$ $(|\psi\rangle \otimes |\varphi\rangle)$ = Tr $(A_1 |\psi\rangle) \cdot$ Tr $(A_2 |\varphi\rangle)$.

If, however, there appear more than one term in the W-representation of the compound system, i.e., if $|c_i| < 1$, then there exist entangled correlations (of the well-known EPR-type) between subsystems $S_1$ and $S_2$. It can be shown in this case that there are no subsystem states $|\xi\rangle$ ($\forall$ $|\xi\rangle \in H_1$) and $|\chi\rangle$ ($\forall$ $|\chi\rangle \in H_2$) such that W is equivalent to the conjoined attribution of $|\xi\rangle$ to subsystem $S_1$ and $|\chi\rangle$ to subsystem $S_2$, i.e., $W \neq |\xi\rangle \otimes |\chi\rangle$.[2] Thus, when a compound system, such as S, is in an entangled state W, namely a superposition of pure states of tensor-product forms, neither subsystem $S_1$ by itself nor subsystem $S_2$ by itself is associated with an individual pure state. The normalised unit vectors $|\psi_i\rangle$, $|\varphi_i\rangle$ belonging to the Hilbert space of either subsystem are not eigenstates of the overall state W. Only the compound system, as a unified whole, is assigned a definite (nonseparable) pure state W, represented appropriately by a state vector in the tensor-product Hilbert space of S. Maximal knowledge of the whole system, therefore, does not allow the possibility of acquiring maximal knowledge of its component parts, a circumstance with no precedence in classical physics. Consequently, the separability principle of Section 1 is violated.

The generic phenomenon of quantum nonseparability casts severe doubts on the existence of *isolated* (sub)systems and the applicability of the notion of atomism, in the sense that the parts of a quantum whole no longer exist as precisely defined individual entities characterized only by intrinsic (non-relational) properties. The nonseparable character of the behavior of an entangled quantum system precludes in a novel way the possibility of describing its component subsystems in terms of pure states. In fact, whenever the pure entangled state of a compound system is decomposed in order to represent subsystems, the effect can only extent up to a representation in terms of statistical (reduced) states of those subsystems. For, whenever a compound system is in an entangled state, as in Eq. (1), there are, in general, no pure states of the component subsystems on the basis of which the compound state of the whole system could be completely determined. Consequently, the legendary notion of the classical paradigm that the nature of the whole is fully describable or reducible to the properties of the parts is no longer defensible. In the framework of quantum mechanics, the state of the whole system cannot in general be determined by the states of its component parts, this being the case even when the parts occupy distinct regions of space however far apart. Since, at the quantum domain, it is exclusively only the compound state of the whole system that exhaustively specifies the probabilistic entangled correlations among the states of its parts. Hence, any case of quantum entanglement constitutes a violation of the separability principle, and the latter is the reason why entanglement induces a sort of holism in quantum mechanics.

The phenomenon of quantum nonseparability indeed reveals the holistic character of entangled quantum systems. Quantum mechanics is the first — and up to day the only — mathematically formulated and empirically well-confirmed theory, which incorporates as its basic feature that the 'whole' is, in a non-trivial way, more than the sum of its 'parts' including their spatio-temporal relations and physical interactions. Contrary to the situation in classical physics, when considering an entangled compound system, 'whole' and 'parts' are related in such a way that their *bi-directional reduction* is, in principle, impossible (see, for instance, Karakostas 2004). Intimately related to this, there exist properties considering any entangled quantum system which, in a clearly specifiable sense, characterize the whole system but are neither reducible to nor implied by or derived from any combination of local properties of its parts. As a means of exemplifying the preceding points, let us consider an important class of compound quantum systems that form the prototype of EPR-entangled systems, namely, spin-singlet pairs. Let then S be a compound system consisting of a pair ($S_1$, $S_2$) of spin-1/2 particles in the singlet state

$$W_S = 1/\sqrt{2} \ \{|\psi_+>_1 \otimes |\varphi_->_2 \ - \ |\psi_->_1 \otimes |\varphi_+>_2\}, \tag{2}$$

where $\{|\psi_\pm >_1\}$ and $\{|\varphi_\pm >_2\}$ are orthonormal bases in the two-dimensional Hilbert spaces $H_1$ and $H_2$ associated with $S_1$ and $S_2$, respectively. As well-known, in such a case, it is quantum mechanically predicted and experimentally confirmed that the spin components of $S_1$ and of $S_2$ have always opposite spin orientations; they are perfectly anticorrelated. Whenever the spin component of $S_1$ along a given direction is found to be $+1/2 \ \hbar$ (correspondingly $-1/2 \ \hbar$), then the spin component of $S_2$ along the same direction must necessarily be found to be $-1/2 \ \hbar$ (correspondingly $+1/2 \ \hbar$), and conversely. From a physical point of view, this derives from the interference (the definite phase interrelations) with which the subsystem states $|\psi\rangle$ and $|\varphi\rangle$ — or, more precisely, the two product states $|\psi_+\rangle_1 \otimes |\varphi_-\rangle_2$, $|\psi_-\rangle_1 \otimes |\varphi_+\rangle_2$ — are combined within $W_S$. This, in turn, leads not only to the subsystem interdependence of the type described above, but also to conservation of the total angular momentum for the pair ($S_1$, $S_2$) of spin-1/2 particles, and thus to the property of definite total spin of value zero for the compound system S.

The latter is a *holistic* property of S: it is not determined by any physical properties of its subsystems $S_1$, $S_2$ considered individually. Specifically, the property of S 'having total spin zero' is not specified by the spin properties of $S_1$ and of $S_2$, since neither $S_1$ nor $S_2$ has any definite spin in the superposed singlet state $W_S$. Moreover, the probability distributions concerning spin components of $S_1$ and of $S_2$ along some one direction do not ensure, with probability one, S's possession of this property. Neither the latter could be understood or accounted for by the possibility (that a strict adherent of reductionism may favor) of treating $S_1$ and $S_2$ separately at the expense of postulating a relation between them as to the effect of their spin components 'being perfectly anticorrelated'. For, while 'having total spin zero' is an intrinsic physical property of the compound system S in the nonseparable state $W_S$, the assumed relation is not an intrinsic physical relation that $S_1$ and $S_2$ may have in and of themselves. That is, although the relation of perfect anticorrelation is encoded within state $W_S$, ascribing this relation to individual parts of a system is not tantamount to being in state $W_S$. The relation of perfect anticorrelation is inherent to the entangled state $W_S$ itself, whose nonseparable nature dictates, in fact, all that can be said about the spin properties of $S_1$ and $S_2$, because it is only the entangled state of the whole that contains the correlations among the spin probability distributions pertaining to the parts.[3] Hence, the part-whole reduction with respect to the property of total spin zero has failed: the latter property, whereas characterizes the whole system, is irreducible to any properties of its constituent parts. Exactly the same holds for the properties of total momentum and relative distance of the overall system S with respect to the corresponding local properties of its parts. Analogous considerations, of course, to the aforementioned paradigmatic case of the spin-singlet pair of particles apply to any case of quantum entanglement. Entanglement need not be of maximal anticorrelation, as in the example of the singlet state. It does neither have to be confined to states of quantum systems of the same kind; entanglement reaches in principle the states of all compound quantum systems.

This is precisely the delicate point with entangled correlations in Hilbert-space quantum mechanics: they cannot be reduced to or explained in terms of pre-assigned relations or interactions among the parts; their existence cannot be traced back to any interactions. Whereas the smallest interaction during the temporal development of the parts of a compound system gives rise to entanglement, entanglement itself needs no recourse to interaction for its being established. Interaction is a sufficient but not a necessary condition for entanglement. Quantum entanglement does occur in the absence of any interactions, since the origin of the phenomenon is essentially of a kinematical rather than dynamical nature. Due to that the entangled correlations among the states of physical systems do not acquire the status of a causally determined relation.[4] Their delineation instead is specified by the entangled quantum state itself which refers directly to the whole system.

To disclose the primary kinematical character of quantum entanglement, as well as its relation to the superposition principle, the states of a system involved in a superposed vector of the form, $|\Psi\rangle = \Sigma_i c_i |\psi_i\rangle$, $\Sigma_i |c_i|^2 = 1$, should be explicitly interpreted so as to refer to *potentially realized* (through the measurement process or 'spontaneously' in nature) states $|\psi_i\rangle$, each possessing a probability amplitude $c_i$ of occurrence. The principle of superposition of states is inseparably linked with the interference of such probability amplitudes, $c_i c_i^*$, reflecting the nature of the *interrelations* among the states of a quantal system. Accordingly, any physical variable *A* that is associated with a superposed state $|\Psi\rangle$ possess *no* definite value at all, rendering unattainable thereby a Boolean yes-no classification of *A* in $|\Psi\rangle$. In other words, for any physical variable *A* in a superposed state $|\Psi\rangle$ of eigenstates of *A* and any proposition *P* concerning *A*, it is not true that either *P* holds or its complement *I–P* holds. In such a circumstance, the possible numerical values of *A* are *objectively indeterminate* and not simply unknown. The objectivity of the indeterminacy of *A* stems from the fact that the probabilities of the various possible outcomes or realizations of *A* are designated by the superposed state itself, a feature with no analogue in classical mechanics.

Consequently, a superposed state or a general quantum-mechanical pure state can be defined independently of measurement only by a probability distribution of potentially possible values which pertain to the physical quantities of a system. Hence, the quantum state may be construed in an ontic sense, regardless of any operational procedures, as representing a network of potentialities, namely, a field of *potentially possible* and not actually existing events.[5] The double modality used here does not simply mean to characterize a transition from potentiality to actuality or from a situation of indefiniteness to definiteness. It also intends to signify that quantum-mechanical potentialities condition but do not control the production of actual events.

The concept of quantum-mechanical potentiality corresponds to the tendency of a quantum system to display a certain measurement result out of a set of multiplied possibilities in case a suitable measurement is made.[6] When, for instance, in the standard EPR-Bohm example, a pair of particles ($S_1$, $S_2$) is in the spin-singlet state of Eq. (2), *no* spin component of either particle exists in a precisely defined form. All three spin components of each particle, however, coexist in a *potential* form and any one component possess the tendency of being actualized at the expense of the indiscriminacy of the others if the associated particle interacts with an appropriate measuring apparatus. As soon as such an interaction takes place, for example, at time $t_o$, and the spin component along a given direction of , say, particle $S_1$ is measured and found equal to $+1/2\hbar$, the subsequent destruction of the superposition bonds (between the tensor-product states involved) imparts to particle $S_2$ a potentiality: that of inducing, with probability one, the opposite value $-1/2\hbar$, if and when, at a time $t > t_o$, particle $S_2$ is submitted to an appropriate measurement of the same component of spin as $S_1$. Thus, the spin-singlet state, furnishing the standard EPR-Bohm example, describes the entanglement, the inseparable correlation of potentialities.

The singlet state (as any entangled state) represents in essence a set of potentialities whose content is not exhausted by a catalogue of actual, pre-existing values that may be assigned to the spin properties of $S_1$ and of $S_2$, separately. It may be worthy to observe in this connection that in the EPR-Bohm example *no* spin property of particle $S_2$ enjoyed a well-defined value *prior* to the measurement performed on $S_1$ and a different value in the sequel. The only change occurring in $S_2$ concerns the transition of a spin property from a potentially possible value to an actually realized value. If the actualized values of all spin properties of $S_2$ were pre-determined, then the behavior of $S_2$ would be fixed in advance regardless of any reference to system $S_1$, and vice versa. Consequently, the behavior of the compound spin-singlet pair would be reducible to the behavior of its parts. A sense of separability would also had been demonstrated if the actualized values of both $S_1$ and $S_2$ were independent or owed their existence to a common cause in the past (e.g., Butterfield 1989). Thus, quantum-mechanical nonseparability is subject to both the actualization of potentialities and the confinement of actualization to the dictates of correlation.

We stress, in this respect, that the concept of quantum-mechanical potentiality should not be classified under an epistemic category of apprehension. It does not refer to someone's deficient knowledge as to the exact nature of a given object, but it belongs to the mode of existence of the object itself.[7] It constitutes, in Aristotelian terminology, "the measure of the actual" (Aristotle, Physics, 207b). It characterizes the degree of realization of a potentially possible event determined by objective physical conditions, namely, by the internal properties of the object and the specified experimental conditions. Quantum-mechanical potentialities are physically real and objective not only in the sense of designating the disposition of an object to expose certain properties when submitted to given conditions, but also in the sense of interfering, under certain circumstances as in quantum coherence or quantum entanglement, with one another.

Thus, when confronting a compound system in an entangled state, one may conceive that the potentialities of subsystems $S_1$ and $S_2$, constituting the whole system, interfere with each other in such a way that the probability of a certain result of a measurement performed on $S_1$ is dependent upon the result of a measurement performed on $S_2$, and conversely.[8] Or even more acutely, as exemplified in the physically important case of maximal spin anticorrelation, the actualization of an arbitrary spin component of $S_1$ entails the actualization of the corresponding spin component of $S_2$ with precisely defined value. Numerous experimental results, whose quantitative character is based on a Bell-type inequality, strongly testify that this aspect of interfering potentialities does take part in nature, even when $S_1$ and $S_2$ are spatially separated (e.g., Aspect et al. 1982; Tittel et al. 1998). Furthermore, as has been repeatedly formally shown, there seems to be no way of utilizing the quantum-mechanical transition from potentiality to actuality for the purpose of establishing a superluminal communication procedure between $S_1$ and $S_2$ that would violate special relativity.[9] Physically, this is in our view due to the fact that the appearance of entangled quantum correlations of events at space-like separation does not refer to the correlation of actual events in external space-time; the creation of entangled correlations occurs in the realm of *potentialities*. While actual events take place in space-time, we never encounter a potentiality at the space-time level. Potential events only refer to the possible existence of physical entities in a space-time region, they do not presuppose the actual presence of physical entities in that region. Hence, potentialities do not materially propagate in external space-time. This also clarifies, in the light of our approach, why entangled correlations cannot be utilized for superluminal communication.

### 3. THE CONTEXT-DEPENDENCE OF QUANTUM OBJECTS AND RELATED PHILOSOPHICAL CONSEQUENCES

From a foundational viewpoint of quantum theory, the concept of quantum entanglement and the associated phenomenon of quantum nonseparability refer to a context-independent, or in d' Espagnat's (1995) scheme, observer- or mind-independent reality. The latter is operationally inaccessible. It pertains to the domain of entangled correlations, potentialities and quantum superpositions obeying a non-Boolean logical structure. Here the notion of an object, whose aspects may result in intersubjective agreement, enjoys no a priori meaning independently of the phenomenon into which is embedded. In quantum mechanics in order to be able to speak meaningfully about an object, to obtain any kind of description, or refer to experimentally accessible facts the underlying wholeness of nature should be decomposed into interacting but disentangled subsystems. As will be argued in the sequel, in consonance with Primas (1993), well-defined separate objects (and their environments) are generated by means of a so-called Heisenberg cut (1958, 116), namely through the process of a deliberate abstraction/projection of the inseparable non-Boolean domain into a Boolean context that necessitates the suppression (or minimization) of entangled correlations between the object-to-be and the environment-to-be (e.g., a measuring apparatus).

The presuppositions of applying a Heisenberg cut are automatically satisfied in classical physics, in conformity with the separability principle of Section 1. In a nonseparable theory like quantum mechanics, however, the concept of the Heisenberg cut acquires the status of a methodological regulative principle through which access to empirical reality is rendered possible. The innovation of the Heisenberg cut, and the associated separation of a quantum object from its environment, is mandatory for the description of measurements (e.g., Atmanspacher 1994). It is, in fact, necessary for the operational account of any directly observable pattern of empirical reality. The very possibility of devising and repeating a controllable experimental procedure presupposes the existence of such a subject-object separation. Without it the concrete world of material facts and data would be inelegible; it would be conceived in a totally entangled manner. In this sense, a physical system may account as an experimental or a measuring device only if it is not holistically correlated or entangled with the object under measurement.

Consequently, any atomic fact or event that 'happens' is raised at the empirical level only in conjunction with the specification of an experimental arrangement [10] — an experimental context that conforms to a Boolean domain of discourse — namely to a set of observables co-measurable by that context. In other words, there cannot be well-defined events in quantum mechanics unless a specific set of co-measurable observables has been singled out for the system-experimental context whole (e.g., Landsman 1995). For, in the quantum domain, one cannot assume, without falling into contradictions, that observed objects enjoy a separate well-defined identity irrespective of any particular context. One cannot assign, in a consistent manner, definite sharp values to all quantum-mechanical observables pertaining to a microscopic object, in particular to pairs of incompatible observables, independently of the measurement context actually specified. In terms of the structural component of quantum theory, this is due to functional relationship constraints that govern the algebra of quantum-mechanical observables, as revealed by the Kochen-Specker (1967) theorem and its recent investigations (e.g., Mermin 1995). In view of them, it is not possible, not even *in principle*, to assign to a quantum system non-contextual properties corresponding to all possible measurements. This means that it is not possible to assign a definite unique answer to every single yes-no question, represented by a projection operator, independent of which subset of mutually commuting projection operators one may consider it to be a member. Hence, by means of a generalized example, if A, B and C denote observables of the same quantum system, so that the corresponding projection operator A commutes with operators B and C ($[A, B] = 0 = [A, C]$), not however the operators B and C with each other ($[B, C] \neq 0$), then the result of a measurement of A *depends* on whether the system had previously been subjected to a measurement of the observable B or a measurement of the observable C or in none of them. Thus, the value of the observable A depends upon the set of mutually commuting observables one may consider it with, that is, the value of A depends upon the set of measurements one may select to perform. In other words, the value of the observable A cannot be thought of as pre-fixed, as being independent of the experimental context actually chosen. In fact, any attempt of simultaneously attributing context-independent, sharp values to all observables of a quantum object forces the quantum statistical distribution of value assignment into the pattern of a classical distribution, thus leading directly to contradictions of the GHZ-type (Greenberger et al. 1990).

This state of affairs reflects most clearly the unreliability of the so-called 'possessed values' principle of classical physics of Section 1, according to which, values of physical quantities are regarded as being possessed by an object independently of any measurement context. The classical-realist underpinning of such an assumption is conclusively shown to be incompatible with the structure of the algebra of quantum-mechanical observables. Well-defined values of quantum observables can, in general, be regarded as pertaining to an object of our interest only within a framework involving the experimental conditions. The latter provide the necessary conditions whereby we make meaningful statements that the properties attributed to quantum objects are part of physical reality. Consequent upon that the exemplification of quantum objects is a *context-dependent* issue with the experimental procedure supplying the physical context for their realization. The introduction of the latter operates as a formative factor on the basis of which a quantum object manifests itself. The classical idealization of sharply individuated objects possessing intrinsic properties and having an independent reality of their own breaks down in the quantum domain. Quantum mechanics describes physical reality in a substantially context-dependent manner.

Accordingly, well-defined quantum objects cannot be conceived of as 'things-in-themselves', as 'absolute' bare particulars of reality, enjoying *intrinsic* individuality or *intertemporal* identity. Instead, they represent carriers of patterns or properties which arise in interaction with their experimental context/environment, or more generally, with the rest of the world;[11] the nature of their existence — in terms of state-property ascription — depends on the context into which they are embedded and on the subsequent abstraction of their entangled correlations with the chosen context of investigation. Thus, the resulting contextual object *is* the quantum object exhibiting a particular property with respect to a certain experimental situation. The contextual character of property-ascription implies, however, that a state-dependent property of a quantum object is not a well-defined property that has been possessed *prior* to the object's entry into an appropriate context. This also means that not all contextual properties can be ascribed to an object at once. One and the *same* quantum object does exhibit several possible contextual manifestations in the sense that it can be assigned several definite incommeasurable properties only with respect to distinct experimental arrangements which mutually exclude each other (for a technical treatment of these elements, see, Section 4). Thus, in contradistinction to a mechanistic or naive realistic perception, we arrive at the following general conception of an object in quantum mechanics. According to this, a quantum object — as far as its state-dependent properties are concerned — constitutes a totality defined by all the possible relations in which this object may be involved. Quantum objects, therefore, are viewed as carriers of inherent dispositional properties. In conjunction with our previous considerations of Section 2, ascribing a property to a quantum object means recognizing this object an *ontic potentiality* to produce effects whenever it is involved in various possible relations to other things or whenever it is embedded within an appropriate experimental context.

Consequently, a quantum object is not an individual entity that possesses well-defined intrinsic properties at all times even beyond measurement interactions, nor is it a well-localized entity in space and time that preserves deterministic causal connections with its previous and subsequent states, allowing it, thereby, to traverse determinate trajectories.[11] In fact, a quantum object exists, independently of any operational procedures, only in the sense of 'potentiality', namely, as being characterized by a set of potentially possible values for its various physical quantities that are actualized when the object is interacting with its environment or a pertinent experimental context. Due to the genuinely nonseparable structure of quantum mechanics and the subsequent context-dependent description of physical reality, a quantum object can produce *no* informational content that may be subjected to experimental testing without the object itself being transformed into a contextual object. Thus, whereas quantum nonseparability refers to an *inner-level* of reality, a mind-independent reality that is operationally elusive, the introduction of a context is related to the *outer-level* of reality, the contextual or empirical reality that results as an abstraction in the human perception through deliberate negligence of the all-pervasive entangled (nonseparable) correlations between objects and their environments (Karakostas 2003). In this sense, quantum mechanics has displaced the verificationist referent of physics from 'mind-independent reality' to 'contextual' or 'empirical reality'.

It should be noted that the concept of a mind-independent reality is not strictly scientific; it does not constitute a matter of physics or mathematics; it is rather metaphysical by nature. It concerns, by definition, the existence of things in themselves regardless of any act of empirical testing. Consequently, it does not apply to empirical science proper. It may be viewed, however, as a regulative principle in physics research, as a conviction which gives direction and motive to the scientific quest. As Einstein put it:

> It is basic for physics that one assumes a real world existing independently from any act of perception. But this we do not *know*. We take it only as a programme in our scientific endeavors. This programme is, of course, prescientific and our ordinary language is already based on it. (quoted in Fine 1986, p. 95)

Granting the metaphysical or heuristic character of its nature, we nonetheless consider the notion of a mind-independent reality as unassailable in any scientific discourse; we amply recognize its existence as being logically *prior* to experience and knowledge; we acknowledge its external to the mind structure as being responsible for *resisting* human attempts in organizing and conceptually representing experience.

But, significantly, in the quantum domain, the nature of this independent reality is left unspecified. For, due to the generalized phenomenon of quantum nonseparability, we must conceive of independent reality as a highly entangled whole with the consequence that it is impossible to conceive of parts of this whole as individual entities, enjoying autonomous existence, each with its own well-defined pure state. Neither reality considered as a whole could be comprehended as the sum of its parts, since the whole, according to considerations of Section 2, cannot be reduced to its constituent parts in conjunction with the spatio-temporal relations among the parts. Quantum nonseparability seems to pose, therefore, a novel limit on the ability of scientific cognizance in revealing the actual character of independent reality itself, in the sense that any detailed description of the latter necessarily results in *irretrievable* loss of information by dissecting the otherwise undissectable. From a fundamental viewpoint of quantum mechanics, any discussion concerning the nature of this indivisible whole is necessarily of an ontological, metaphysical kind, the only confirmatory element about it being the network of entangled interrelations which connect its events. In this respect, it can safely be asserted that reality thought of as a whole is not scientifically completely knowable, or, at best, in d' Espagnat's (1995) expression, it is veiled. Hence, our knowledge claims to reality can only be partial, not total or complete, extending up to the structural features of reality that are approachable by the penetrating power of the theory itself and its future development.

The term 'reality' in the quantum realm cannot be considered to be determined by what physical objects really are in themselves. As already argued, this state of affairs is intimately associated with the fact that, in contrast to classical physics, values of quantum-mechanical quantities cannot, in general, be attributed to a quantum object as intrinsic properties. The assigned values cannot be said to belong to the observed object alone regardless of the overall experimental context which is relevant in any particular situation. Hence, well-defined quantum objects, instead of picturing entities populating the mind-independent reality, they depict the *possible manifestations* of these entities within a concrete experimental context (see, in particular, Section 4). In this respect, the quantum-mechanical framework seems only to allow a detailed description of reality that is co-determined by the specification of a measurement context. Without prior information of the kind of observables used to specify a context and thus to prepare a quantum-mechanical state, it is just not possible to find out what the actual state of a quantum system is; measurement of observables that do not commute with this original set will inevitably produce a different state. What contemporary physics, especially quantum mechanics, can be expected therefore to describe is not 'how mind-independent reality is', as classical physics may permit one to presume. Within the domain of quantum mechanics, knowledge of 'reality in itself', 'the real such as it truly is' independent of the way it is contextualized, is impossible in principle.[13] Thus, it is no longer conceivable to judge the reliability of our knowledge through a comparison with reality itself, and in the scientific description we must adopt alternative necessary conditions for meeting a suitable criterion of objectivity.

To this end we underline the fact that although contextual objects cannot be viewed, by definition, as objects in an absolute, intrinsic sense, nonetheless, they preserve *scientific objectivity*; they reflect structures of reality in a manner that is independent of various observers or of any observer's cognition. For, since they are given at the expense of quantum-mechanical nonseparability, the 'conditions of their being experienced' are determined by the 'conditions of accessibility', or more preferably, in reinterpreting Cassirer (1936/1956, p. 179) in the above expression, by the '*conditions of disentanglement*'. Once the latter conditions are specified, the result of their reference is intersubjective, hence mind-independent. In other words, given a particular experimental context, concrete objects (structures of reality) have well-defined properties independently of our knowledge of them. Thus, within the framework of quantum mechanics, the perceptible separability and determinateness of the contextual objects of empirical reality are generated by means of an experimental intervention that suppresses (or sufficiently minimizes) the factually existing entangled correlations of the object concerned with its environment. It is then justified to say that the fulfillment of disentanglement conditions provides a level of description to which one can associate a separable, albeit contextual, concept of reality whose elements are commonly experienced as distinct, well-localized objects having determinate properties.

Furthermore, since the contextual object constitutes the actually verifiable appearance of the quantum object, quantum objects are *objectively real* in the sense that they are manifested to us in the context of lawful connections; they also contribute to the creation of such lawful connections. Hence, we are confronted in the quantum domain with a reversal of the classical relationship between the concepts of object and law, a situation that has been more vividly expressed in broader terms (of a neo-kantian type, not necessarily adopted here in toto) by Cassirer. In his words:

> ... objectivity itself — following the critical analysis and interpretation of this concept — is only another label for the validity of certain connective relations that have to be ascertained separately and examined in terms of their structure. The tasks of the criticism of knowledge ("Erkenntniskritik") is to work backwards from the unity of the general object concept to the manifold of the *necessary and sufficient conditions that constitute it*. In this sense, that which knowledge calls its "object" breaks down into *a web of relations* that are held together in themselves through the highest rules and principles. (Cassirer 1913, transl. in Ihmig 1999, p. 522; emphasis added)

Although Cassirer's reference is within the context of relativity theory, where these 'highest rules and principles' stand for the symmetry principles and transformations which leave the relevant physical quantities invariant, in the quantum domain, a precondition of something to be viewed as an object of scientific experience is the elimination of the entangled correlations with its environment. In other words, in order for any object-system S of the quantum realm, its observed qualities (e.g., any obtainable measuring results on S) to be considered as properties of S, the condition of disentanglement must be fulfilled. Thus, disentanglement furnishes a necessary condition for a quantum object to become amenable to scientific analysis and experimental investigation; that is, disentanglement constitutes a necessary material precondition of quantum physical experience by rendering the object system S a scientific object of experience.

## 4. CONTEXTUALITY AS A DISENTANGLEMENT PREPARATION PROCEDURE

We explicitly point out that the fulfillment of such disentanglement conditions is provided by contextuality, namely, the pre-selection of a suitable experimental context on which the state of a measured system can be conditioned. How then one may proceed in purely formal quantum-mechanical terms so as to establish this fact and also recover our previous general conception of a quantum object? To begin with, we give a known important auxiliary result concerning the objectification or actual occurrence of an observable (or property) in a quantum state of an object. If W is a general initial state of a quantum object S, and if we consider a change of state W→$W_A$ satisfying the requirement that an observable A of S acquires a well-defined value a in $W_A$ — i.e., the corresponding characteristic projector $P_a$ satisfies the relation Tr $W_A P_a$ = 1 — then necessarily we have

$$W_A = \Sigma_i P_a^{(i)} W P_a^{(i)}, \quad i = 1, 2, \ldots . \tag{3}$$

Evidently, Eq. (3) is the Lüders (1951) formula in the so-called 'non sorting' version where no information is extracted as to the specific measurement result $a_i$ of A. If the initial density operator W is an idempotent density operator, namely, a one-dimensional projection operator W = |ψ><ψ| representing a pure quantum state |ψ> = $\Sigma_i c_i |a_i>$, and the spectrum of the observable A is assumed to be non-degenerate, then Eq. (3) is restricted to

$$W_A = \Sigma_i |c_i|^2 |a_i><a_i|, \quad i = 1, 2, \ldots . \tag{4}$$

In the framework of our considerations, the transition W→$W_A$ should not be considered as the result of a reduction or decoherence process, although, of course, in quantum measurement interactions such effects may actually occur. Thus, the state $W_A$ is not regarded here as the final post-measurement state of our object of interest, but, on the contrary, as a *conditionalization preparation procedure* of the *initial* state of the object with respect to the measurement context of observable A. In this sense, the transition W→$W_A$ seems to offer the possibility of implementing a contextual-realist interpretation of the quantum-mechanical formalism at the level of states. Hence, the contextual state $W_A$ may naturally be viewed to represent microscopic reality in the context of a measurement of a particular observable A.

As argued in the preceding section, this move is physically justified by the fact that quantum-mechanical measurements are capable of only probing contextual states. It cannot be overemphasized that in the absence of the specification of a measurement context, the customary concept of a general pure quantum state is not directly amenable to experimental investigation. In view of the fact that the projection operators, or, equivalently, the corresponding one-dimensional closed linear subspaces, can be used to represent both the pure states of a given object and also its quantum-mechanical propositions of yes-no experimental tests, let us consider the following proposition 'p' stating that 'the state vector associated with the projection operator W = |ψ><ψ| lies in the subspace $H_W$ of the object's relevant Hilbert state-space' (i.e., 'W∈ $H_W$ ⊆ H'). Certainly, proposition 'p' determines, up to a phase factor, a pure state of the object. But does proposition 'p' state an empirical fact? Does 'p' itself describe a quantum-mechanical event? The answer is unequivocally no. For, as already underlined, the occurrence of any particular kind of a quantum-mechanical event implies the use of a particular set of observables, that is, it presupposes the specification of a definite set of empirical predicates. The subspaces $H_W$ of the Hilbert state-space, however, neither represent nor determine specific empirical predicates for quantum-mechanical events. The latter correspond with the axes of possible parameterizations of $H_W$, i.e., with an orthogonal basis of eigenvectors of a selected observable to be measured, or, more generally, with an orthogonal basis of joint eigenvectors of a set of mutually compatible observables for $H_W$.

Now, given that a particular observable has been selected and hence a particular basis in $H_W$ has been chosen, the occurrence of a particular quantum-mechanical event presupposes also the existence of an appropriate measurement context relative to which the measuring conditions remain invariant. Formally, a measurement context $M_A$ can be defined by a pair (W, A), where, as in our preceding considerations, W = |ψ><ψ| is an idempotent projection operator denoting the general initial state of a system and A = $\sum_i a_i P_i$ is a self-adjoint operator denoting the measured observable. Then, following Howard (1994) (see also Fine 1987), we say that a contextual state, mathematically defined by a density operator $W_A$, is *representative* for $M_A$ — when A is a non-degenerate observable — just in case Eq. (4) is satisfied, namely, $W_A = \sum_{i=1}^{n} |c_i|^2 |a_i><a_i|$, where each $|a_i>$ is an eigenvector of A and $|c_i| = |<ψ, a_i>|$, i = 1, ..., n. In other words, $W_A$ is a mixed state over a set of basis states that are eigenstates of the measured observable A, and it reproduces the probability distribution that W assigns to the values of A. Thus, with respect to the representative contextual state $W_A$ the following conditions are satisfied:

i) Each |a_i> is an eigenvector of A. Thus, each quantum-mechanical proposition $P_{|a_i>}$ = |a_i><a_i|, i = 1, ..., n, assigns in relation to $M_A$ some well-defined value to A (i.e., the eigenvalue $α_i$ satisfying A|a_i> = $α_i$|a_i>).

ii) Any eigenvectors |a_i>, |a_j>, i≠j, of A are orthogonal. Thus, the various possible propositions {$P_{|a_i>}$}, i = 1, ..., n, are mutually exclusive within $M_A$. In this sense, the different orthogonal eigenstates {|a_i>}, i = 1, ..., n, correspond to different results of the measured observable A or to different settings of the apparatus situated in the context $M_A$.

iii) Each |a_i> is non-orthogonal to W=|ψ><ψ|. Thus, each proposition $P_{|a_i>}$ whose truth value is not predicted with certainty is possible with respect to $M_A$.

It is evident, therefore, that the contextual state $W_A$ represents the set of all probabilities of events corresponding to quantum-mechanical propositions $P_{|a_i>}$ that are associated exclusively with the measurement context $M_A$. In it the propositions $P_{|a_i>}$ correspond in a one-to-one manner with disjoint subsets of the spectrum of the observable A and hence generate a Boolean lattice of propositions (e.g., Jauch 1968, pp. 97-101). Thus, the $P_{|a_i>}$–propositions are assigned determinate truth values, in the standard Kolmogorov sense, by the state $W_A$.

Furthermore, the description in terms of the representative contextual state $W_A$ is essentially equivalent for predictive purposes to the description in terms of the state W, since, according to the Hilbert-space formalism of quantum mechanics, for any observable A, TrWA = Tr$W_A$A. The latter equality designates the fact that the initial general state W does not have any observational significance over the contextual state as long as predictions of measurement outcomes are restricted to the observables that are actually measured within the context of applicability of $W_A$. In other words, with respect to the specification of a measurement context $M_A$, the contextual state description $W_A$ does yield the same results as the general state description W. Yet, this equivalence is a consequence of the quantum-mechanical framework itself. Of course, observational distinguishability between the two descriptions appears whenever one considers observables incompatible to the representative (selected) observable that determines a particular measurement context. However, if A and B are incompatible observables, then the state $W_B$ cannot be prepared in the measurement context $M_A$ of a measurement of A, because $M_A$≠$M_B$. Incompatible observables acquire distinct contextual states, since, due to complementarity, they are physically well established in interaction to distinct, mutually exclusive, measurement arrangements. Contextuality is at the basis of quantum-mechanical complementarity. Hence, given a general initial state of an object, *identical* preparation of contextual states referring to the object's incompatible observables are impossible in quantum mechanics due to mutual exclusiveness of measurement arrangements that incommeasurability implies.[14] Consequently, the observational equivalence between W and a representative contextual state $W_R$, associated to an arbitrary observable R, is both meaningful and permissible only as a context by context equivalence, where the appropriate context is defined each time by the selection of the measured observable R.

In this sense, contextual quantum states represent probability distributions for the results of measurements from the point of view of the measurement context in which they are given. In connection to our preceding considerations of the 'potentia' conception, we may naturally say that the probability distribution defined by $W_A$ represents the potentially possible values for the measured observable A within the context $M_A$. Or, to put it another way, whereas the initial general state W may be thought to represent a network of potentialities pertaining to a quantum object independently of any operational act, $W_A$ represents the potentialities available to the object after a preparation procedure has been set up to measure A. Hence, contextual quantum states do not represent the potentialities inherent in the object as such, independently of the *kind* of measurement actually to be performed. They represent instead *Boolean perspectives* of such potentialities as viewed from the measurement context in which they are expressed. Thus, in conformity with our previous conception of a quantum object, the totality of all those possible perspectives, corresponding to a complete set of mutually exclusive measurement contexts that are available to an object, exhausts the information content that may be extracted from the object concerned.[15] In this manner, the quantum object of scientific experience may be thought of as the whole class of its aspects, 'seen' from mutually different contexts.

We finally note in concluding this work that, due to the aforementioned statistical observational equivalence $TrWA = TrW_AA$ of Hilbert-space quantum mechanics, all information provided by any quantal measurement procedure, with respect to a particular measurement context $M_A$, is consistent with the view that this information may indeed refer to the representative contextual state $W_A$ rather than to the state W the microscopic object was in before its embedding into $M_A$. To explicate this point in a physically unambiguous manner, it is crucial to re-emphasize that the contextual state $W_A$ is not the post-measurement state, reached after an A-measurement has been carried out on the object concerned. It is neither a reduced statistical state resulting from the mathematical operation of tracing out certain degrees of freedom from the compound object-plus-apparatus system. The contextual state represents here an alternative description of the initial state W of the object system S; it constitutes a state preparation of S in the context of the observable actually measured. The transition $W \rightarrow W_A$ to a contextual state, therefore, need not be regarded as the result of any physical procedures or actions which, when carried out, may change a particular physical object. The transition, in other words, must not be viewed as a material physical process imposed on the object under measurement, but as a *redescription* of the measured object which is necessitated by taking specifically into account the context of the selected observable. For, it is important to realize that this kind of redescription is intimately related to the fact that both states W and $W_A$ represent the *same* physical object S, albeit in different ways. Whereas W refers to a general initial state of S independently of the specification of any particular observable, and hence, regardless of the determination of any measurement context, the state $W_A$ constitutes an initial state preparation of S that is adapted to the observable actually being measured, while dropping all 'irrelevant' reference to observables that are incompatible with such a preparation procedure. And, as we saw, with respect to the observable determining the kind of preparation, namely the measurement context, the states of the measured object and the measuring apparatus can be *separated* or *disentangled*, thus allowing us to give a consistent interpretation of their statistics that corresponds to well-defined facts. Moreover, such a separating procedure is fully consistent with the unitary dynamical evolution of quantum-mechanical systems. In this regard, the aforementioned crucial distinction between the two kinds of initial states seems to be also significant in re-assessing the notorious quantum measurement problem from a new perspective. An elaborate development of this state of affairs, however, is left to a future work.

## NOTES

[1] In this work we shall make no detailed reference to alternative interpretations of ordinary quantum mechanics as, for instance, Bohm's ontological or causal interpretation.

[2] In a paper related to the Einstein-Podolsky-Rosen argument, Schrödinger remarked with respect to this distinctive feature of nonfactorizability as follows: ''When two systems, of which we know the states by their respective representations, enter into temporary physical interaction due to known forces between

them, and then after a time of mutual influence the systems separate again, then they can no longer be described in the same way as before, viz. by endowing each of them with a representative of its own. ... I would not call that one but rather *the* characteristic trait of quantum mechanics, the one that enforces its entire departure from classical lines of thought'' (Schrödinger 1935/1983, p. 161).

[3] In this connection see Esfeld (2004). Also Rovelli (1996) and Mermin (1998) highlight the significance of correlations as compared to that of correlata.

[4] For instance, the entangled correlations between spatially separated systems can not be explained by assuming a direct causal influence between the correlated events or even by presupposing the existence of a probabilistic common cause among them in Reichenbach's sense. Butterfield (1989) and van Fraassen (1989) have shown that such assumptions lead to Bell's inequality, whereas, as well known, the latter is violated by quantum mechanics. See in addition, however, Belnap and Szabó (1996) for the notion of a *common* common cause in relation to quantum correlations occurring in the Greenberger-Horne-Zeilinger (GHZ) situation.

[5] The following account of Heisenberg is characteristic of the significance of the notion of potentiality as a clarifying interpretative concept for quantum theory. He writes: "The probability function ... contains statements about possibilities or better tendencies ('potentia' in Aristotelian philosophy), and these statements ... do not depend on any observer. ... The physicists then speak of a 'pure case'. ... [Consequently,] it is no longer the objective events but rather the probabilities for the occurrence of certain events that can be stated in mathematical formulae. It is no longer the actual happening itself but rather the possibility of its happening — the potentia — that is subject to strict natural laws. ... If we ... assume that even in the future exact science will include in its foundation the concept of probability or possibility — the notion of potentia — then a number of problems from the philosophy of earlier ages appear in a new light, and conversely, the understanding of quantum theory can be deepened by a study of these earlier approaches to the question" (Heisenberg 1958, p. 53; 1974, pp. 16, 17).

[6] See, for instance, Heisenberg (1958, pp. 42, 185); Popper (1980, Ch. 9; 1990, Ch. 1); Shimony (1993, Vol. 2, Ch. 11). Margenau (1950, pp. 335-337, 452-454) has also used the concept of 'latency' to characterize the indefinite quantities of a quantum-mechanical state that take on specified values when an act of measurement forces them out of indetermination.

[7] It should be analogously underlined that the pure quantum-mechanical state does not constitute a mere expression of our knowledge of a particular microphysical situation, as frequently nowadays construed, thus acquiring an epistemic status (e.g., Fuchs and Peres 2000). The quantum state designates an economic and effective embodiment of *all possible manifestations* of the quantum-mechanical potentialities pertaining to a system; it encapsulates all facts *concerning the behavior* of the system; it codifies not only what may be 'actual' in relation to the system, upon specified experimental conditions, but also what may be 'potentially possible', although it does not literally represent any concrete features of the system itself. The latter element, however, does not abolish the character of quantum state as a bearer of *empirical content*, since the quantum state modulates as a whole the characteristics or statistical distributions of realizable events. Furthermore, the statistical distribution of any such event — associated to a certain preparation procedure of

a quantum state — is a stable, reproducible feature. Thus, the *generator* of these statistical regularities, namely, the quantum state associated with preparation, can be considered as an *objective* feature of the world (for details, see Sections 3 and 4).

[8] The probabilistic dependence of measurement outcomes between spatially separated systems forming an entangled quantum whole corresponds, as an expression of the violation of the separability principle, to the violation of what has been coined in the Bell-literature as Jarrett's (1984) 'completeness condition', or equivalently, Shimony's (1986) 'outcome independence' condition. A detailed description of these conditions would fall outside the scope of the present work. A review of them may be found in Howard (1997).

[9] A recent generalized version of the so-called no-signalling theorem is given by Scherer and Busch (1993).

[10] It should be pointed out that Bohr already on the basis of his complementarity principle introduced the concept of a 'quantum phenomenon' to refer "exclusively to observations obtained under specified circumstances, including an account of the whole experiment" (Bohr 1963, p. 73). This feature of context-dependence is also present in Bohm's ontological interpretation of quantum theory by clearly putting forward that "quantum properties cannot be said to belong to the observed system alone and, more generally, that such properties have no meaning apart from the total context which is relevant in any particular situation. In this sense, this includes the overall experimental arrangement so that we can say that measurement is context dependent" (Bohm and Hiley 1993, p. 108).

[11] Note that the so-called invariant or state-independent, and therefore, context-independent properties — like 'rest-mass', 'charge' and 'spin' — of elementary objects-systems can only characterize a certain class of objects; they can only specify a certain sort of particles, e.g., electrons, protons, neutrons, etc. They are not sufficient, however, for determining a member of the class as an individual object, distinct from other members within the same class, that is, from other objects having the same state-independent properties. Thus, an 'electron', for instance, could not be of the particle-kind of 'electrons' without fixed, state-independent properties of 'mass' and 'spin', but these in no way suffice for distinguishing it from other similar particles or for 'individuating' it in any particular physical situation. For a detailed treatment of this point, see, for example, Castellani (1999).

[12] In standard quantum mechanics, it is not possible to establish a causal connection between a property A(t) at time t and the same property A(t´´) at a later time t´´, both pertaining to an object-system S, if S had been subjected at a time value t´, t<t´<t´´, to a measurement of a property B incompatible with A. Because the successive measurement of any incompatible property of this kind would provide an uncontrollable material change of the state of S. Thus, a complete causal determination of all possible properties of a quantum object, most notably, coordinates of position and their conjugate momenta, allowing the object, henceforth, to traverse well-defined trajectories in space-time is not possible.

[13] It is tempting to think that a similar sort of context-dependence already arises in relativity theory. For instance, if we attempt to make context-independent attributions of simultaneity to spatially distant events — where the context is now determined by the observer's frame of reference — then we will come into conflict with the experimental record. However, given the relativization of simultaneity — or the relativization of properties like length, time duration, mass, etc. — to a reference frame of motion, there is

nothing in relativity theory that precludes a complete description of the way nature is. Within the domain of relativity theory, the whole of physical reality can be described from the viewpoint of any reference frame, whereas, in quantum mechanics such a description is inherently incomplete.

[14] It is worthy to note in this association that the impossibility of preparing the same initial contextual state in mutually exclusive measurement arrangements constitutes a sufficient condition for preventing derivability of the Bell inequality without invoking nonlocality (e.g., De Baere 1996).

[15] It has been shown by Ivanovic (1981), Wooters and Fields (1989), see also, Brukner and Zeilinger (1999), that the total information content of a quantum system represented by a density matrix (pure or mixed) is optimally obtainable from a complete set of mutually exclusive (complementary) measurements corresponding to the system's complete set of mutually complementary observables.

REFERENCES

Aspect, A., Grainger, G., and Roger, G.: 1982, 'Experimental Test of Bell's Inequalities Using Time-Varying Analyzers', *Physical Review Letters* **49**, 1804-1807.

Atmanspacher, H.: 1994, 'Objectification as an Endo-Exo Transition', in H. Atmanspacher and G.J. Dalenoort (eds.), *Inside Versus Outside*, Berlin, Springer, pp. 15-32.

Baere, W. De.: 1996, 'Quantum Nonreproducibility at the Individual Level as a Resolution of Quantum Paradoxes', in A. Mann and M. Revzen (eds.), *The Dilemma of Einstein, Podolsky and Rosen – 60 Years Later*, Bristol, Institute of Physics Publishing, pp. 95-108.

Belnap, N. and Szabó, L. E.: 1996, 'Branching Space-time Analysis of the GHZ Theorem', *Foundations of Physics* **26**, 989-1002.

Blank, J., Exner, P., and Havlicek, M.: 1994, *Hilbert Space Operators in Quantum Physics*, New York, American Institute of Physics.

Bohm, D., and Hiley, B.: 1993, *The Undivided Universe. An Ontological Interpretation of Quantum Theory*, London, Routledge.

Bohr, N.: 1963, *Essays 1958-1962 on Atomic Physics and Human Knowledge*, New York, Wiley.

Brucner, C., and Zeilinger, R.: 1999, 'Operationally Invariant Information in Quantum Measurements', *Physical Review Letters* **83**, 3354-3357.

Butterfield, J.: 1989, 'A Space-Time Approach to the Bell Inequality', in J. Cushing and E. McMullin (eds.), *Philosophical Consequences of Quantum Theory: Reflections on Bell's Theorem*, Notre Dame, University of Notre Dame Press, pp. 114-144.

Castellani, E.: 1999, 'Galilean Particles: An example of Constitution of Objects', in E. Castellani (ed.), *Interpreting Bodies*, Princeton, Princeton University Press, pp. 181-196.

Cassirer, E.: 1936/1956, *Determinism and Indeterminism in Modern Physics*, Yale University Press.

Einstein, A.: 1948/1971, 'Quantum Mechanics and Reality', in M. Born (ed.), *The Born-Einstein Letters*, London, Macmillan, pp. 168-173.

Esfeld, M.: 2004, 'Quantum Entanglement and a Metaphysics of Relations', *Studies in the History and Philosophy of Modern Physics* **35**, 601-617.

Espagnat, B. De.: 1995, *Veiled Reality*, Reading, Addison-Wesley.

Fine, A.: 1986, *The Shaky Game*: *Einstein, Realism and the Quantum Theory*, Chicago, University of Chicago Press.

Fine, A.: 1987, 'With Complacency or Concern: Solving the Quantum Measurement Problem', in R. Kargon and P. Achinstein (eds.), *Kelvin's Baltimore Lectures and Modern Theoretical Physics*, Cambridge: MA, MIT Press, pp. 491-506.

Fuchs, C., and Peres, A.: 2000, 'Quantum Theory Needs No Interpretation', *Physics Today* **3**, 70-71.

Greenberger, D., Horn, M., Shimony, A., and Zeilinger, A.: 1990, 'Bell's Theorem Without Inequalities', *American Journal of Physics* **58**, 1131-1143.

Healey, R.: 1991, 'Holism and Nonseparability', *The Journal of Philosophy* **LXXXVIII**, 393-421.

Heisenberg, W.: 1958, *Physics and Philosophy*, New York, Harper & Row.

Heisenberg, W.: 1974, *Across the Frontiers*, New York, Harper & Row.

Howard, D.: 1989, 'Holism, Separability and the Metaphysical Implications of the Bell Experiments', in J. Cushing and E. McMullin (eds.), *Philosophical Consequences of Quantum Theory*: *Reflections on Bell's Theorem*, Notre Dame, University of Notre Dame Press, pp. 224-253.

Howard, D.: 1994, 'What Makes a Classical Concept Classical?', in J. Faye and H. Folse (eds.), *Niels Bohr and Contemporary Philosophy*, New York, Kluwer, pp. 201-229.

Howard, D.: 1997, 'Space-Time and Separability: Problems of Identity and Individuation in Fundamental Physics', in R. Cohen, M. Horne and J. Stachel (eds.), *Potentiality, Entanglement and Passion-at-a Distance*, Dordrecht, Kluwer, pp. 113-141.

Ihmig, K.: 1999, 'Ernst Cassirer and the Structural Conception of Objects in Modern Science: The Importance of the "Erlanger Programm"', *Science in Context* **12**, 513-529.

Ivanovic, I.D.: 1981, 'Geometrical Description of Quantal State Determination', *Journal of Physics A* **14**, 3241-3245.

Jarrett, J.P.: 1984, 'On the Physical Significance of the Locality Conditions in the Bell Arguments', *Noûs* **18**, 569-589.

Jauch, J.: 1968, *Foundations of Quantum Mechanics*, Reading: MA, Addison-Wesley.

Karakostas, V.: 2003, 'The Nature of Physical Reality in the Light of Quantum Nonseparability', *12th International Congress of Logic, Methodology and Philosophy of Science*, Oviedo, Volume of Abstracts, 329-330.

Karakostas, V.: 2004, 'Forms of Quantum Nonseparability and Related Philosophical Consequences', *Journal for General Philosophy of Science* **35**, 283-312.

Kochen, S. and Specker, E.: 1967, 'The Problem of Hidden Variables in Quantum Mechanics', *Journal of Mathematics and Mechanics* **17**, 59-87.

Landsman, N.: 1995, 'Observation and Superselection in Quantum Mechanics', *Studies in History and Philosophy of Modern Physics* **26**, 45-73.

Luders, G.: 1951, 'Ueber die Zustandsänderung Durch den Messprozess', *Annalen der Physik* **8**, 322-328.

Margenau, H.: 1950, *The Nature of Physical Reality*, New York, McGraw Hill.

Mermin, N. D.: 1995, 'Limits to Quantum Mechanics as a Source of Magic Tricks: Retrodiction and the Bell-Kochen-Specker Theorem', *Physical Review Letters* **74**, 831-834.

Mernin, N. D.: 1998, 'What is Quantum Mechanics Trying to Tell Us?', *American Journal of Physics* **66**, 753-767.

Popper, K.R.: 1980, *The Logic of Scientific Discovery*, London, Hutchinson.

Popper, K.R.: 1990, *A World of Propensities*, Bristol, Thoemmes.

Primas, H.: 1993, 'The Cartesian Cut, the Heisenberg Cut, and Disentangled Observers', in K.V. Laurikainen and C. Montonen (eds.), *Symposia on the Foundations of Modern Physics*, Singapore, World Scientific, pp. 245-269.

Rovelli, C.: 1996, 'Relational Quantum Mechanics', *International Journal of Theoretical Physics* **35**, 1637-1678.

Scheibe, E.: 1973, *The Logical Analysis of Quantum Mechanics*, Oxford, Pergamon Press.

Scherer, H., and Busch, P.: 1993, 'Problem of Signal Transmission via Quantum Correlations and Einstein Incompleteness in Quantum Mechanics', *Physical Review A* **47**, 1647-1651.

Shimony, A.: 1986, 'Events and Processes in the Quantum World', in R. Penrose and C. Isham (eds.) *Quantum Concepts in Space and Time*, Oxford, Oxford University Press, pp. 182-203.

Shimony, A.:1993, *Search for a Naturalistic World View*, Volume 2, *Natural Science and Metaphysics*, Cambridge, Cambridge University Press.

Schrödinger, E.: 1935/1983: 'The Present Situation in Quantum Mechanics', *Naturwissenschaften* **22**, 807-812, 823-828, 844-849. Reprinted in J. Wheeler and W. Zurek (eds.), *Quantum Theory and Measurement*, Princeton, Princeton University Press, pp. 152-167.

Tittel, W., Brendel, J., Zbinden, H., and Gisin, N.: 1998, 'Violation of Bell Inequalities by Photons More Than 10km Apart', *Physical Review Letters* **81**, 3563-3566.

Van Fraassen, B. C.: 1989, 'The Charybdis of Realism: Epistemological Implications of Bell's Inequality', in J. Cushing and E. McMullin (eds.), *Philosophical Consequences of Quantum The*ory: *Reflections on Bell's Theorem*, Notre Dame, University of Notre Dame Press, pp. 97-113.

Wooters, W., and Fields, B.: 1989, 'Optimal State-Determination by Mutually Unbiased Measurements', *Annals of Physics* **191**, 363-381.